\documentclass[reprint,
 amsmath,amssymb,
 aps
]{revtex4-2}
\usepackage{hyperref}
\usepackage{xcolor,soul}
\usepackage{mathtools}
\usepackage{verbatim}
\usepackage{graphicx}
\usepackage{dcolumn}
\usepackage{bm}
\makeatletter
\newsavebox{\@brx}
\newcommand{\llangle}[1][]{\savebox{\@brx}{\(\m@th{#1\langle}\)}%
  \mathopen{\copy\@brx\kern-0.5\wd\@brx\usebox{\@brx}}}
\newcommand{\rrangle}[1][]{\savebox{\@brx}{\(\m@th{#1\rangle}\)}%
  \mathclose{\copy\@brx\kern-0.5\wd\@brx\usebox{\@brx}}}
\makeatother
\usepackage{stmaryrd}
\usepackage{braket}
\usepackage{enumerate}
\usepackage{enumitem}
\usepackage{simpler-wick}
\usepackage[normalem]{ulem}

\usepackage{lmodern}
\usepackage{microtype}

\DeclareFontEncoding{U}{}{}
\DeclareFontFamily{U}{bbold}{}
\DeclareFontShape{U}{bbold}{m}{n}
 {  <5> <6> <7> <8> <9> gen * bbold
   <10> <10.95> bbold10
  <12> <14.4> bbold12
 <17.28> <20.74> <24.88> bbold17
  }{}
\DeclareSymbolFont{bbold}{U}{bbold}{m}{n}
\DeclareSymbolFontAlphabet{\mathbbold}{bbold}

\newcommand{\id}{\mathbbold{1}}
\def\cZ{\mathcal{Z}}
\def\MWK{{\rm MWK}}
\def\st{{\rm string}}
\def\tbar{\overline{t}}
\def\ccrit{\mathcal{C}_{\rm crit}}
\def\sfa{\mathsf{a}}
\def\sfb{\mathsf{b}}

\def\RMT{{\rm RMT}}

\usepackage{physics}

\def\llangle{\langle\!\langle}
\def\rrangle{\rangle\!\rangle}

\def\sl{SL(2,\Z)}

\def\s{\sigma}

\def\L{\Lambda}

\def\z{\zeta}

\def\cF{{\cal F}}

\def\cM{{\cal M}}

\def\cO{{\cal O}}

\def\cZ{{\cal Z}}

\def\Re{{\rm Re \,}}
\def\Im{{\rm Im \,}}

\def\Tr{{\rm Tr}}

\def\half{{1\over 2}}
\def\p{\partial}

\def\g{\gamma}
\def\G{\Gamma}
\def\l{\lambda}
\def\eps{\epsilon}

\def\sl{SL(2,\Z)}
\def\crit{{\half+i\w}}

\def\cF{\mathcal{F}}
\def\w{\omega}
\def\L{\Lambda}
\def\half{{1\o2}}

\def\R{\mathbb{R}}
\def\qb{\overline{q}}

\def\Re{\text{Re}}
\def\Im{\text{Im}}

\def\x{\times}

\def\eps{\epsilon}

\def\hb{\overline h}
\def\Z{\mathbb{Z}}

\def\1{{\rm 1-loop}}

\def\cZ{\mathcal{Z}}

\def\Tr{{\rm Tr}}

\def\c{\cite}

\def\cM{\mathcal{M}}

\def\c{\cite}

\def\G{\Gamma}
\def\p{\partial}
\def\o{\over}
\def\g{\gamma}
\def\D{\Delta}
\def\rar{\rightarrow}

\def\eqr{\eqref}

\def\O{{\cal O}}

\def\ssec{\subsection}
\def\sssec{\subsubsection}
\def\sec{\section}
\def\i{\infty}
\def\foot{\footnote}
\newcommand{\es}[2] {\begin{equation} \label{#1} \begin{split} #2 \end{split} \end{equation}}
\newcommand{\e}[2] {\begin{equation} \label{#1} #2 \end{equation}}

\newcommand{\beq}{\begin{equation}}
\newcommand{\eeq}{\end{equation}}
\newcommand{\beqy} {\begin{eqnarray}}
\newcommand{\eeqy} {\end{eqnarray}}
\newcommand{\bsmat}{\begin{smallmatrix}}
\newcommand{\esmat}{\end{smallmatrix}}
\newcommand{\bmat}{\begin{matrix}}
\newcommand{\emat}{\end{matrix}}

\def\({\left(}
\def\){\right)}
\def\[{\left[}
\def\]{\right]}

\def\<{\langle}
\def\>{\rangle}

\def\g{\gamma}
\def\G{\Gamma}

\def\D{\Delta}
\def\z{\zeta}

\def\l{\lambda}

\def\t{\tau}
\def\s{\sigma}

\begin{document}


\title{AdS$_3$ Pure Gravity and Stringy Unitarity}

\author{Gabriele Di Ubaldo, Eric Perlmutter
}

\affiliation{\small Universit\'e Paris-Saclay, CNRS, CEA, Institut de Physique Th\'eorique, 91191, Gif-sur-Yvette, France
}


\begin{abstract}

We construct a unitary, modular-invariant torus partition function of a two-dimensional conformal field theory with a Virasoro primary spectral gap of $\Delta_* = \frac{c-1}{12}$ above the vacuum. The twist gap is identical, apart from two states $\mathcal{O}_*$ with spin scaling linearly in the central charge $c$. These states admit an AdS$_3$ interpretation as strongly coupled strings. All other states are black hole microstates.

\end{abstract}

\maketitle

The quest for AdS$_3$ pure gravity still beckons.

It is not fully known whether, or in what precise sense, a consistent such theory exists, either quantum mechanically or in the semiclassical limit. The latter is of particular physical interest, due to the existence of black holes and the emergence of spacetime. 

Holographically speaking, the outstanding spectral problem is to find a torus partition function of a two-dimensional conformal field theory (CFT) that is mutually compatible with unitarity (a non-negative Virasoro primary spectral density) and modularity (exact $SL(2,\Z)$-invariance of the partition function), while preserving the spectral gaps of a dual bulk theory with only black holes above a normalizable AdS$_3$ ground state. No known partition function satisfies these basic requirements. 

There exists a diverse set of approaches to this problem which, famous as it is, we describe in condensed fashion. Summing over all smooth on-shell 3-manifolds $\cM$ with $\p \cM = T^2$ \c{Maloney:2007ud}, namely the $\sl$ family of BTZ black holes, generates a negative density of states in two regimes \c{Maloney:2007ud,Keller:2014xba,Benjamin:2019stq}: at large spin $j\rar\i$ near extremality, 
\e{eq:MWKneg1}{\int_0^{t_0}dt \,\rho_{{\rm MWK},\,j}(t)\sim  (-1)^j e^{\pi \sqrt{\xi j}} , \quad t_0\sim e^{-2\pi \sqrt{\xi j}}}
where 
\e{}{t := \text{min}(h,\hb) - \xi\,, \quad j=h-\hb\,,\quad \xi:=\frac{c-1}{24}\,,}
 and at the scalar black hole threshold,
\e{eq:MWKneg2}{\rho_{{\rm MWK},0}(t)= -6\delta(t) + (t>0~\text{continuum})\,.}
The property \eqr{eq:MWKneg1} is especially severe: an exponentially large negative density despite an exponentially small window. From the bulk perspective, seeking a consistent pure gravity path integral requires reckoning with the sum over topologies; for related work, see \cite{Dijkgraaf:2000fq,Manschot:2007ha,Yin:2007gv,Castro:2011zq,Cotler:2020ugk, Eberhardt:2020bgq, Eberhardt:2021jvj, Eberhardt:2022wlc}. (We note here some recent work in AdS$_3$/CFT$_2$ that studies fixed bulk topologies \cite{Belin:2020hea,Chandra:2022bqq, Collier:2023fwi, Yan:2023rjh, Chandra:2023rhx, Abajian:2023bqv}.)

Some valuable progress has been made. Explicit restoration of unitarity may be achieved by retreating from pure gravity \cite{Alday:2019vdr,Benjamin:2020mfz}, adding heavy point-particle matter which admits a geometric bulk interpretation. The construction of \cite{Maxfield:2020ale}, which preserves the pure gravity spectrum, uses dimensional reduction to JT gravity to fix \eqr{eq:MWKneg1} with an infinite sum over off-shell Seifert manifolds, though it remains a mostly \cite{DiUbaldo:2023qli} implicit construction away from extremality and leaves \eqr{eq:MWKneg2} intact. Other approaches that forego a subset of the above conditions include \c{Cotler:2018zff,Li:2019mwb,Mertens:2022ujr}.

\sec{Partition function}

The Virasoro primary partition function is defined as
\e{}{Z_p(\t) = \sqrt{y} |\eta(\t)|^2 Z(\t)}
where $Z(\t) = \Tr_{\mathcal{H}}(q^{L_0 -{c\o 24}}\qb^{\bar{L}_0-{c\o 24}})$ is the  torus partition function (non-holomorphic) and $\t:=x+iy$. The following modular-invariant $Z_p(\t)$ is unitary at sufficiently large $\xi$:
\e{zmain}{\cZ(\t) = Z_{\rm MWK}(\t) + Z_{\rm string}(\t) }
where
\es{Zdefs}{Z_{\rm MWK}(\t) &:= \sum_{\g\in SL(2,\Z)/\Gamma_\i} \sqrt{\Im(\g\t)} \,|q_\g^{-\xi}(1-q_\g)|^2\\
Z_{\rm string}(\t) &:= \sum_{\g\in SL(2,\Z)/\Gamma_\i} \sqrt{\Im(\g\t)} \,\(2q_\g^{\xi/4} \qb_\g^{-\xi/4} + \text{c.c.}\)}
with $q_\g := e^{2\pi i \g \t}$. These are Poincar\'e sums over $\sl$ modulo $\G_\i$, the set of modular $T$-transformations \foot{We thus formally demand that $\xi\in 2\Z_+$, though we note that $\cZ(\t)$ remains unitary if we shift the seed quantum numbers while preserving $T$-invariance. In particular, $\cZ(\t)$ remains unitary if we slightly shift $\D_* > 2\xi$ while keeping $t_* = -{\xi\o4}$ fixed; for simplicity, we focus on the case $\D_*=2\xi$ in the main text.}. As we substantiate below, the unitary range of $\xi$ includes $\xi\gg1$, and provisionally appears to hold for all $\xi\in 2\Z_+$. The reason for the ``string'' moniker will be explained momentarily. 

From a CFT point of view, $Z_\st(\t)$ is a Poincar\'e sum over two copies of a Virasoro primary seed state $\O_*$ with quantum numbers
\e{qns}{(\D_*,j_*) = \(2\xi,{\xi\o2}\)\quad \Leftrightarrow \quad (t_*,\tbar_*) = \(-{\xi\o4},{\xi\o4}\)}
and its parity image with $h_* \leftrightarrow \hb_*$. We have employed the ``reduced twist'' variable $t$ along with its partner $\tbar := \text{max}(h,\hb) - \xi$. We have chosen the state in \eqr{Zdefs} to be doubly-degenerate, a natural choice that preserves integrality, but $\cZ(\t)$ is unitary for a finite range of degeneracies $d_*>1$ (see Appendix \ref{App:positivity}, e.g. Fig. \ref{fig:den73}). 

\begin{figure}[t]
\centering
{
\includegraphics[scale=0.4]{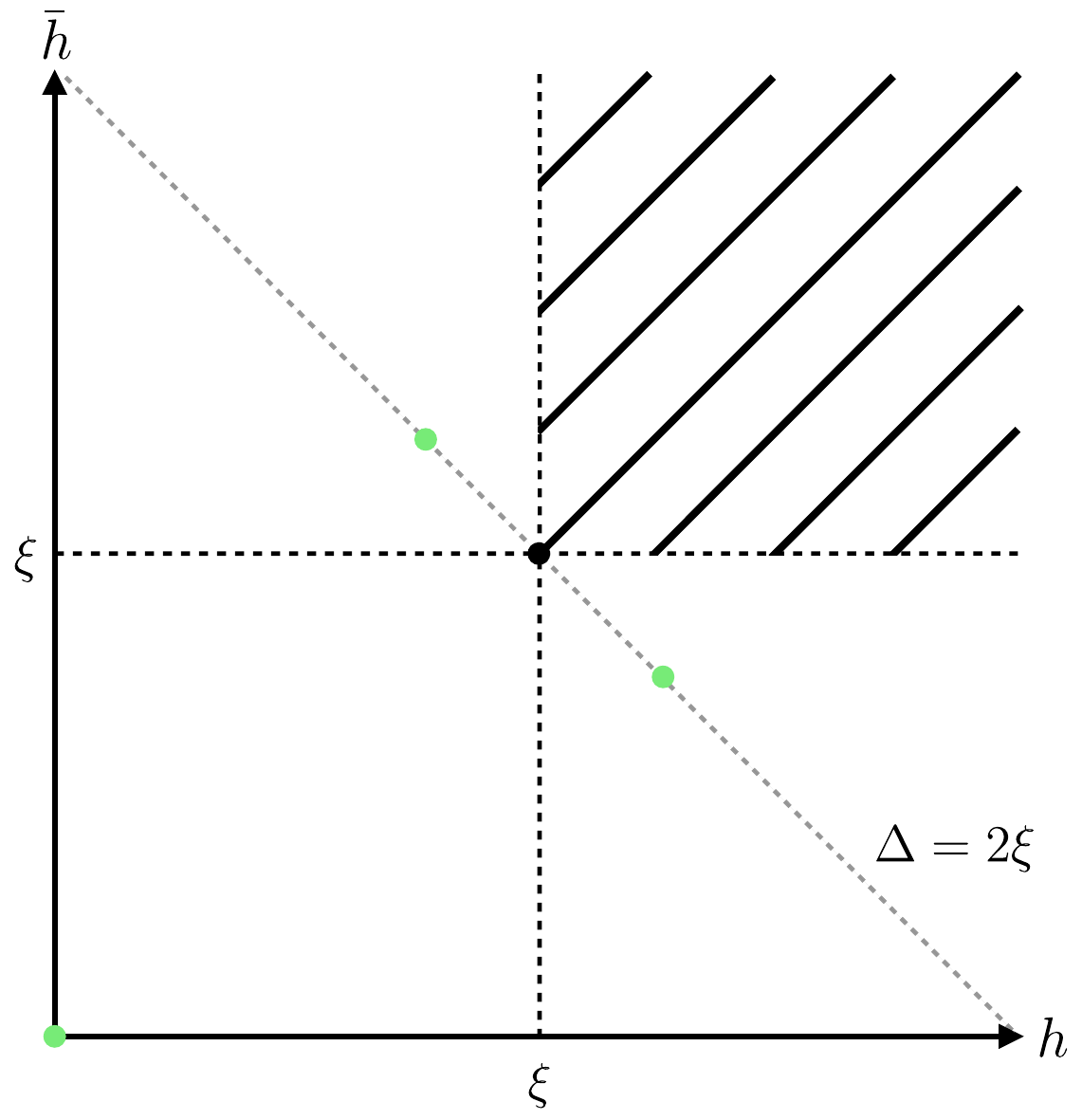}
}
\caption{The Virasoro primary spectrum of $\cZ(\t)$. Green dots denote the vacuum state and two (parity-invariant) states with $(h_*,\hb_*)=(\frac{5}{4}\xi,\frac{3}{4}\xi)$, interpretable in AdS$_3$ as strongly coupled strings. All other states exceed the semiclassical black hole threshold: min$(h,\hb)\geq \xi$. The density of states is positive.}
\vspace{.1in}
\label{fig:spectrum}
\end{figure}

Let us state the spectral properties of the partition function $\cZ(\t)$, deferring its unitarity to the next subsection.  The spectrum is shown in Fig. \ref{fig:spectrum}. The gap in conformal dimension above the vacuum is exactly
\e{Dgap}{\Delta_{*}=\frac{c-1}{12}}
with no corrections. This is the value anticipated by the Virasoro modular bootstrap program (e.g. \cite{Hellerman:2009bu,Friedan:2013cba,Collier:2016cls,Afkhami-Jeddi:2019zci,Hartman:2019pcd}) as the optimal gap at {\it large} $c$, on the basis of black hole universality: the conformal dimension \eqr{Dgap} corresponds to the massless limit of the semiclassical BTZ black hole. The state-of-the-art bootstrap upper bound on the spectral gap at large $c$ is the numerical result \cite{Afkhami-Jeddi:2019zci}
\e{}{\Delta_{*} \lesssim  \frac{c}{9.08} \qquad (c\gg1)}
with a slightly weaker analytical bound \c{Hartman:2019pcd}. (See \c{Hellerman:2010qd, Keller:2012mr, Qualls:2013eha, Hartman:2014oaa, Fitzpatrick:2014vua, Kim:2015oca, Benjamin:2016fhe, Benjamin:2016aww, Collier:2017shs, Anous:2018hjh, Collier:2018exn, Benjamin:2018kre, Kusuki:2018wpa, Maxfield:2019hdt,Ganguly:2019ksp, Mukhametzhanov:2019pzy, Benjamin:2020zbs, Pal:2020wwd, Kaidi:2020ecu, Lin:2021bcp, Pal:2023cgk} for further bootstrap work on Virasoro spectra at large $c$.) The explicit realization by $\cZ(\t)$ of the gap \eqr{Dgap} while preserving unitarity at $\xi \gg 1$ (the first such example, to our knowledge) is also noteworthy because of the paucity of pure CFT arguments that a gap this large is possible. Conversely, $\cZ(\t)$ shows constructively that without incorporating discreteness into the modular bootstrap \c{Kaidi:2020ecu}, the optimal bound on the gap cannot be lower than $\Delta_*=2\xi$. This statement applies for all values of $\xi$ for which $\cZ(\t)$ is unitary. 

As for the twist spectrum, all Virasoro primaries besides $\id$ and $\O_*$ have $t\geq 0$. There is a positive integer number of scalar states at $t=0$ (see \eqr{rhoscal}). The spectrum of $t>0$ states is continuous. 
This can be understood rather generally in terms of coarse-graining. At large $\xi$, this can be thought of as a consequence of ignorance of exponentially small effects in $c$ -- for example, smearing over the mean level spacing $\sim e^{-S_{{\rm Cardy},j}(t)}$. We explain these points of interpretation in Sec. \ref{sec:random}.

\ssec{Density of states}

The corresponding Virasoro primary density of states, related to our partition function as
\e{}{{\cZ(\t)\o \sqrt{y}} = \sum_{j=0}^\i (2-\delta_{j,0}) \cos(2\pi j x) \int_\mathbb{R} d\D\,e^{-2\pi y(\D-2\xi)}\rho_j(\D)}
can be derived straightforwardly using existing methods for Poincar\'e sums. We have, in terms of reduced twist $t$,
\e{eq:rhotot}{\rho_j(t)=\rho_{{\rm MWK},\,j}(t)+ \rho_{{\rm string},\,j}(t)}
for every spin $j$. The MWK density $\rho_{{\rm MWK},\,j}(t)$ is recalled in \eqr{eq:rhomwk}. The new term is, for $j\neq 0$, 
\es{eq:rhostring}{\rho_{{\rm string},\,j}(t) &= \frac{4}{\sqrt{t\,\tbar}}\sum_{s=1}^{\infty} f_{j,j_*;s}\cos\Big(\frac{2\pi}{s}\sqrt{\xi \tbar}\Big)\cosh\Big(\frac{2\pi}{s}\sqrt{\xi t}\Big)\\&+(j_* \rar -j_*, t \leftrightarrow \tbar)}
where $f_{j,j_*;s} : = {S(j,j_*;s)/s}$ with $S(j,j_*;s)$ a Kloosterman sum \eqr{kldef}. For $j=0$, such sums must be regularized; using standard methods nicely summarized in \c{Benjamin:2020mfz}, the result is the $j=0$ specialization of the $j\neq 0$ densities, augmented by a constant subtraction; see \eqr{eq:rhomwkreg} and \eqr{eq:rhostringreg}. 

There are two hurdles to establishing positivity: one must cancel the negativity of the MWK partition function in the $j\rar\i$ regime, and at the scalar threshold $t=0$, both without introducing new negativity. 

At $j\rar\i$, the negativity \eqr{eq:MWKneg1} is resolved by construction: we have added states with reduced twist $t_* = -\xi/4$, designed precisely to avoid the large-spin negativity in accordance with the arguments of \cite{Benjamin:2019stq} and the subsequent approach of \cite{Alday:2019vdr,Benjamin:2020mfz}. (We added two such states, but any number $d_*>1$ would do; we review this in Appendix \ref{App:positivity}.) The states $\O_*$ have asymptotically large spin as $\xi\rar\i$. It is exactly this property which admits the novelty of a spectral gap $\D_* = 2\xi$ without introducing further negativity elsewhere in the spectrum -- and indeed, as we now show, curing the scalar negativity \eqr{eq:MWKneg2} in the process.

The scalar density of states is
\e{rhoscal}{\rho_0(t)=\delta(t+\xi) + (-6+8\sigma_0(j_*))\delta(t) + \tilde{\rho}_{0}(t)\,.}
The first term is the vacuum state. The second, formerly problematic, term has been rendered strictly positive, for any $j_* = \xi/2$.  Happily, it is also an integer, a welcome surprise. Unlike previous approaches to this negativity, its resolution does not require the addition of an ``extra'' ad hoc $+6\delta(t)$ \cite{Keller:2014xba}, instead coming for free in $\rho_{\st,0}(t)$. We note a number-theoretic feature of this degeneracy: if $j_*$ is prime, then $\sigma_0(j_*)=2$.

The last term, $\tilde{\rho}_{0}(t)$, is the continuum with support on $t>0$, given explicitly in \eqr{eq:regtot}.  Its positivity requires a more careful analysis because various large-$\xi$ suppression factors are absent when $j=0$, i.e. $t = \tbar$, as can be seen in \eqr{eq:rhostring}; however, $\tilde\rho_0(t)$ is indeed positive for all $t\geq 0$. We provide details in Appendix \ref{App:positivity}, but can sketch the essential point here. In the regime $\xi t\gg1$, the scalar MWK density is positive and exponentially larger in magnitude than the string density. As $\xi t\sim \O(1)$, positivity is non-trivial as both densities are of the same order and the string density is term-wise oscillatory in $t$. With an eye toward semiclassical gravity, we focus on $\xi \gg1$, taking the regime of $x:= 2\pi\sqrt{\xi t}$ fixed.  The proof of positivity proceeds in two steps: first, we show that the sum of the $s=1$ and $s=2$ terms in \eqr{eq:regtot} is positive; and second, we show that the  $s\geq 2$ terms are individually positive. 

\begin{figure}[t]
\centering
{
\includegraphics[scale=0.67]{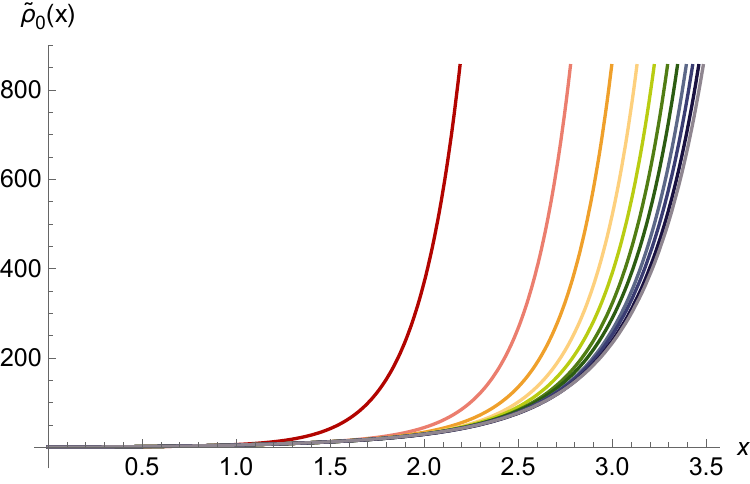}
}
\caption{Plot of the regularized scalar primary density of states $\tilde{\rho}_{0}(x)$ of the partition function $\cZ(\t)$, as a function of $x=2\pi \sqrt{\xi t}$, with $\xi$ ranging from $\xi=2$ (red) to $\xi=102$ (blue) in steps of 10. The curves are positive for all $x\geq 0$. (Obtained by summing over $s \leq 200$ in \eqr{eq:regtot}.)}
\vspace{.1in}
\label{fig:densityd2}
\end{figure}

Numerical evaluation of the scalar density at large but finite $\xi$ confirms these analytic results, as shown in Fig. \ref{fig:densityd2}. Indeed, we see that positivity appears to hold all the way down to $j_*=1$, i.e. $\xi=2$, formally the smallest central charge in our construction \foot{One can generalize our construction to the case of $j_*\in\Z_+-\half$, whereupon the stabilizer group $\G_\i$ in \eqr{Zdefs} would instead be the set of $T^2$ transformations. Analogous extensions to supersymmetric partition functions would also be straightforward. It would be interesting to ask whether the requisite positivity is satisfied in each of these cases. See \cite{Maloney:2007ud,Keller:2012mr,Bae:2018qym,Benjamin:2016aww,Benjamin:2020zbs} for related material.}. 

\sec{A bulk string interpretation}

The above construction is purely on the CFT side. Is there an AdS$_3$ gravity interpretation of the highly-spinning operator $\O_*$ and its modular images?

One appealing answer is that $\O_*$ is a strongly coupled string, and its modular images, stringy contributions to the black hole spectrum. While an operator like $\O_*$ with $t<0$ and $\tbar >0$ cannot be dual to a smooth BTZ black hole nor to a conical defect (such solutions with real mass and angular momenta do not exist in semiclassical AdS$_3$ gravity coupled to point particles), spinning strings in AdS$_3$ can, and indeed do, satisfy this condition. 

The spectrum of folded, spinning Nambu-Goto strings coupled to gravity in AdS$_3$ was studied in \cite{Maxfield:2022rry} in the classical limit. The Virasoro primary string spectrum is parameterized by a string tension $\l$ and an angular velocity $\w$. The string tension is given in terms of AdS, string and Planck scales as 
\e{tension}{\l = {1\o 2\pi}{L_{\rm AdS}\o \ell_s}{\ell_p\o\ell_s}}
where $\ell_p = 8\pi G_N$. 

For a given $\l$, the string spectrum starts at the origin $t=\tbar = -\xi$ and ends at the extremality bound $t=0$ or $\tbar=0$. Matching the string spectrum to the quantum numbers \eqr{qns} of the operator $\O_*$ yields the unique result
\e{l1w2}{\l_*=1\,,\quad \w_*=2\,.}
This string is strongly coupled: from \eqr{tension}, an AdS-sized string with $\l_*=1$ requires $\ell_p/\ell_s \sim \O(1)$, which is the ratio that defines an effective string coupling $g_s = (\ell_p/\ell_s)^{>0}$ (where the exponent depends on the details of the putative string background \foot{For example, in the D1-D5 system $\l \sim g_6/\sqrt{c}$ where $g_6$ is the six-dimensional effective string coupling, so $\l=1$ is highly quantum. In general, $\l\sim \O(1)$ captures a ``very strongly coupled'' limit of fixed coupling $g_s$, rather than a 't Hooft-type double-scaling limit. One can see this by rewriting $\l = 6(L/\ell_s)^2/c$ using the Brown-Henneaux relation \cite{Brown:1986nw}.}). 

The specific value $\l_*=1$ happens to enjoy a certain synergy with the equations of \cite{Maxfield:2022rry}. For generic $\l$ and $\w$, the solutions of \cite{Maxfield:2022rry} are given in terms of elliptic integrals. However, at $\l=1$ -- and only at $\l=1$ -- the solution simplifies dramatically, as the string embedding equation becomes algebraic. It is simple enough to recall explicitly in a few lines. The Lorentzian spacetime metric outside the string is locally AdS$_3$ with the corresponding mass and angular momentum,
\e{}{ds^2 = {1\o 16}(-du^2 + dv^2) - \(z-{1\o 256z}\) du dv + {dz^2 \o 4z^2}}
where, in the conventions of \cite{Maxfield:2022rry}, the conformal boundary is at $z\rar\i$. The string embedding is determined by functions $u(\s,\t), v(\s,\t)$ and $z(\s)$, where $(\s,\t)$ are worldsheet coordinates with induced metric
\es{}{h &= \Omega^2(\s)(-d\t^2 + d\s^2)\\
\Omega^2(\s) &= 3{(z_L-z(\s))(z(\s) - z_R)\o z(\s)}}
where \eqr{l1w2} implies $z_L = {3\o 16}, z_R = -{1\o 48}$, and 
\e{}{z(\s) = \frac{\left(32-25 \cos^2\s+5 \sqrt{25 \cos ^4\s-64 \cos ^2\s+64}\right)}{384} }
Opposite points on the string are identified, ``sewing up'' the spacetime \foot{This identification does not happen at fixed $t$, so the spacetime does not ``pinch off'' on a constant time slice.}. The outermost radius of the string (where it folds) is at $z = z_L$, while the center of the string is at $z=z(0) = {1/ 12}$. The spacetime ends at the string, avoiding a naked singularity. 

So we see that the state $\O_*$ admits an interpretation as a highly-spinning string coupled to gravity in AdS$_3$. That it is strongly coupled dovetails nicely with how AdS$_3$ pure gravity could possibly arise in string theory: strong coupling is necessary to gap the light string modes to the Planck scale.

\ssec{Black hole microstates}

Our construction adds not only the states $\O_*$, but their $\sl$ images too. These states are heavy, but are {\it not} BTZ black holes (fully captured by the MWK sum over smooth Euclidean saddles) nor their orbifolds (which are modular images of conical defect geometries). 

Instead, these are new black hole microstates made of strongly coupled strings. The Euclideanized, modular-transformed solutions of \cite{Maxfield:2022rry} are small black strings, in the following specific sense: whereas a BTZ/conical defect solution with the same quantum numbers would be nakedly singular, the strings shroud this region by terminating the spacetime. These geometries may be thought of as quantum AdS$_3$ versions of the stringy cloak of \cite{Dabholkar:2004dq} and other small black strings (e.g. \c{Sen:2004dp,Castro:2008ne}). That they are ``small'' -- the modular transforms of a single string, rather than a parametrically large number of them -- is also visible thermodynamically in the different functional forms of the stringy and BTZ microcanonical entropies: $\rho_{\st,j}(t)$ is oscillatory as a function of $t$, unlike the BTZ density $\rho_{\MWK,j}(t)$, and is exponentially subleading to $\rho_{\MWK,j}(t)$, term-by-term in the modular sum, away from the near-extremal regime $\xi t \lesssim \O(1)$ where the BTZ black hole becomes highly quantum \foot{This is within the ``enigmatic'' regime \cite{Hartman:2014oaa} where such corrections may be large, consistent with modularity and sparseness.}. This fluctuating behavior signals that the stringy degrees of freedom give genuinely new contributions to the black hole Hilbert space, distinct from the semiclassical BTZ geometries or quotients thereof.

\sec{$SL(2,\mathbb{Z})$ spectral representation}

As a slight detour, it is enlightening to give an alternative representation of $Z_\st(\t)$. The spectral gap condition $\D_* = 2\xi$ implies that $Z_\st(\t)\in L^2(\cF)$, and hence admits a harmonic decomposition in the $SL(2,\Z)$ spectral eigenbasis, comprised of the completed Eisenstein series $E_\crit^*(\t)$ with $\w\in\R$ and Maass cusp forms $\phi_n(\t)$ (e.g. \cite{Terras_2013,Benjamin:2021ygh,Collier:2022emf}). Denoting their spin-$j$ Fourier coefficients as $\sfa^{(s)}_{j}$ and $\sfb^{(n)}_{j}$, respectively, and using the conventions of \cite{DiUbaldo:2023qli}, we have
\es{zstspec}{Z_{\rm string}(\t) &= \int_{\ccrit} \,\sfa^{(s)}_{j_*}\, {\Gamma \left(\frac{\half-s}{2}\right) \Gamma \left(\frac{s-\half}{2}\right)\o \L(s)\L(1-s)}E^*_{s}(\t)\\&~~\,+ \sum_{n=1}^\i \sfb^{(n)}_{j_*} \Gamma \Big(-{i \w_n\o2}\Big) \Gamma \Big(\frac{i \w_n}{2}\Big) \phi_n(\t)}
where $\ccrit$ denotes ($(4\pi i)^{-1}$ times) contour integration along $s=\half+i\w$, and $\L(s) := \pi^{-s} \G(s)\z(2s)$ is the completed Riemann zeta function. (See Appendix \ref{app:zspec}.) 

Presenting $Z_\st(\t)$ in spectral form reveals some interesting features and curiosities.

First, the modular average of $Z_\st(\t)$ vanishes:
\e{}{\<Z_{\rm string}\> := \int_\cF {dx dy\o y^2} Z_\st(\t) =  0\,.}
This follows from the vanishing of the Eisenstein spectral overlap in \eqr{zstspec} at $s=0$, which defines the modular average in general.  We note that this property is shared by Narain CFTs \cite{Benjamin:2021ygh}.  

Next, $Z_\st(\t)$ may be written as the action of an $\sl$ Hecke operator $T_{\xi/2}$ \foot{Acting on non-holomorphic $SL(2,\Z)$-invariant functions, $T_j f(\t) =  {1\o \sqrt{j}} \sum_{a,b,d} f\qty({a\t+b\o d})$ where $ad = j,\, d>0$ and $0\leq b \leq d-1$.} on a ``primitive'' partition function, $\mathcal{Z}_\st(\t)$, defined as $Z_\st(\t)$ but with the Fourier coefficients evaluated at $j_*=1$:
\e{}{\cZ(\t) = Z_{\rm MWK}(\t) + T_{\xi/2} \,\mathcal{Z}_\st(\t)\,.} 
In this way, the entire family of unitary partition functions indexed by $\xi$ may be generated by a Hecke action, implementing shifts in central charge. This shares a superficial likeness with Witten's construction of holomorphic extremal CFT partition functions \cite{Witten:2007kt}, with obvious differences. 

Finally, there is a profound conjecture in number theory, the ``horizontal'' Sato-Tate conjecture for Maass cusp forms of $\sl$, which has interesting consequences for the spectral decomposition \cite{sarnak,hejhal,Steil:1994ue,Then_2004}. The conjecture states that for prime $j \rar\i$ and any fixed $n$, the normalized Fourier coefficients $\sfb^{(n)}_{j}/\sfb^{(n)}_1$ are equidistributed with respect to Wigner's semicircle distribution. This (and $\sfb^{(n)}_1 \neq 0$, which follows from Hecke relations applied to Hecke-Maass cusp forms) implies that
\e{}{\lim_{j\rar\i}\llangle{\sfb^{(n)}_{j}}\rrangle = 0 \qquad (\text{fixed $n$})}
where $\llangle \cdot \rrangle$ indicates a statistical average. Therefore, even though $(Z_\st,\phi_n) \propto \sfb^{(n)}_{j_*} \neq 0$, they vanish on average in the large central charge limit $j_*\rar\i$ \foot{The conjecture is widely held to be true on the basis of stringent numerical checks, limited proofs, and relation to Sato-Tate conjectures in other contexts. Note that at finite $j_*$, the ``vertical'' Sato-Tate conjecture \cite{Sarnak_1987}, would only imply vanishing coefficients as $n\rar\i$.}. In this sense, the Eisenstein term seems to more directly underlie the unitarity of $\cZ(\t)$. It would be nice to understand this from a physical, quantum chaos point of view.

\sec{Summary and Random (matrix) comments}\label{sec:random}

Our main result is the construction of the unitary partition function $\cZ(\t)$ given in \eqr{zmain}, with the spectral gaps depicted in Fig. \ref{fig:spectrum}. 

From the AdS$_3$ gravity point of view, despite the dimension gap above the vacuum state to the black hole threshold $\D_* = {c-1\o 12}$, this is not a semiclassical pure gravity path integral in the strict sense, due to the spinning states $\O_*$ with sub-threshold twist. At any finite spin, these states are not visible, and the theory contains only black hole states. The degeneracies of all discrete states are integers. 

We have advanced a bulk interpretation of $\O_*$ as a strongly coupled spinning string, though other interpretations may well be possible (or preferred). We view this as an indicative toy model for a genuine string theory compactification to AdS$_3$ pure gravity. A complete approach would include higher Regge trajectories; corrections to the spectrum from excitations around the spin-$j$ ground states of \cite{Maxfield:2022rry}; and the other ingredients, such as fluxes and their brane sources, required to solve the strongly coupled string field equations (whatever they may be).

\ssec{Randomness}

Our construction cures the negativity from the sum over smooth bulk saddles semiclassically, rather than quantum mechanically. Quantum effects are not just present in a consistent theory, but are expected to be crucial in the engineering of a bona fide theory of AdS$_3$ pure gravity: there are strong indications that if such a theory exists, off-shell geometries encoding random matrix behavior of the chaotic spectrum play a central role in unitarizing the spectrum \cite{Saad:2019lba,Cotler:2020ugk,Maxfield:2020ale}. An explicit determination of the leading-order random matrix contribution to the semiclassical path integral of pure gravity with torus boundary, denoted as $Z_\RMT(\t)$, was made in \cite{DiUbaldo:2023qli}. 

In any theory of semiclassical AdS$_3$ gravity (pure or otherwise), the black hole spectrum is chaotic, and its path integral should encode random fluctuations for quantum consistency. Such random matrix contributions are absent in $\cZ(\t)$. We may explain this fact, as well as the continuous spectrum in the chaotic regime $t>0$, by interpreting $\cZ(\t)$ as the partition function of a microscopic compact CFT that has been subject to coarse-graining. 

As shown in \cite{DiUbaldo:2023qli} using a formalism built on a CFT trace formula, the random matrix contribution to the density of states, properly understood, vanishes upon coarse-graining the spectrum over a suitable microcanonical window in twist, $\delta t$ \foot{This coarse-graining may be performed using convolution as $\overline{f(t)} := \int_0^\i dt'\,W(t-t') f(t')$ where $W(t)$ has characteristic width $\delta t$ and obeys $\int_0^\i dt\,W(t) = 1$.}. Because this window is necessarily larger than the exponentially small mean level spacing of the chaotic spectrum, the coarse-graining simultaneously explains both the absence of random matrix contributions to \eqr{zmain} and its continuous spectrum while remaining compatible with a microscopic CFT interpretation. Given our explicit construction, we can determine $\delta t$: it is the characteristic wavelength of the oscillations of $\rho_{{\rm string},j}(t)$ in \eqr{eq:rhostring}, namely, $\delta t\sim 1/\xi$. We emphasize that this coarse-graining interpretation does not rely on a $\xi \gg 1$ limit, and is compatible with compactness of a putative underlying CFT; there could, of course, be as-yet-unknown bootstrap constraints that rule this possibility out. 

Note that in a $\xi \gg 1$ limit, $\cZ(\t)$ is also compatible with other interpretations, in particular with a hypothetical ensemble average over (possibly near-)CFTs, or with other, perhaps independent, constructions of ``approximate CFT'' \c{Schlenker:2022dyo}. While we have presented a microscopic CFT interpretation in part to emphasize that a departure from standard AdS/CFT physics is not required at this level, semiclassical gravity seems unable to distinguish among these \cite{Pollack:2020gfa,Schlenker:2022dyo,Cotler:2022rud}, at least perturbatively in $G_N$.

A complementary view on this coarse-grained interpretation comes from the formalism of \cite{DiUbaldo:2023qli}. Since $Z_\st(\t)$ is the modular completion of a non-black hole state, we do not expect it to encode random matrix behavior per se \cite{Cotler:2016fpe,Schlenker:2022dyo,DiUbaldo:2023qli, Haehl:2023tkr}. Applying the results of \cite{DiUbaldo:2023qli} to $Z_\st(\t)$ helps to ratify this perspective. In \eqr{zstspec} we provided the $\sl$ spectral decomposition of $Z_\st(\t)$. A canonical diagnostic of random matrix universality is the presence of a linear ramp in the coarse-grained spectral form factor, with a specific coefficient prescribed by the random matrix ensemble. We can ask whether $Z_\st(\t)$ generates this ramp after squaring and taking the diagonal approximation. A necessary and sufficient condition for the ramp was derived in \cite{DiUbaldo:2023qli}, as an exponential decay condition on the spectral overlaps at $\w\rar\i$. One readily checks that $Z_\st(\t)$ does not satisfy this criterion, instead decaying as a power law \foot{Similarly, the Hecke projection \cite{DiUbaldo:2023qli} of $Z_{\st}(\t_1)Z_{\st}(\t_2)$ is not a wormhole amplitude.}. 

\ssec{Stringiness}
On the other hand, $Z_\st(\t)$ exhibits some behavior that lies somewhere ``in between'' chaotic and non-chaotic. Define a microcanonical coarse-graining over mean twist,
\e{}{\overline{f(t_1)f(t_2)} := \int_0^\i dt'\, f(t'+\eps)f(t'-\eps) W(t-t')}
where $t={t_1+t_2\o 2}$ and $\eps = {t_1-t_2\o 2}$. Applying this to $f(t) = \rho_{{\rm string},j}(t)$ at fixed $j$ using \eqr{eq:rhostring} produces a non-zero variance upon coarse-graining over windows $\delta t \gtrsim \frac{1}{\xi}$. However, its oscillatory behavior leads to suppression relative to the disconnected average. In particular, at $\xi\gg1$, 
\e{}{\frac{\text{Var}(\rho_j(t))}{\bar{\rho}_j(t)^2}\approx e^{-4\pi \sqrt{\xi(t+j)}}\qquad (\xi\gg1)}
where $\bar{\rho}_j(t)=\rho_{{\rm MWK},j}(t)$. In the extremal limit $t\rar 0$, the suppression factor is $e^{-S_{0,j}}$, where $S_{0,j}= 4\pi \sqrt{\xi j}$ is the extremal spin-$j$ BTZ black hole entropy. In contrast, wormholes encoding chaotic behavior are suppressed as $e^{-2S_{0,j}}$ in the extremal limit \cite{Saad:2018bqo,Ghosh:2019rcj,Saad:2019lba,Cotler:2020ugk,DiUbaldo:2023qli}. It would be worthwhile to understand this intermediate behavior as a non-perturbative effect, possibly associated to strongly coupled strings, in a UV complete AdS$_3$ gravity path integral.

\sec*{Acknowledgments}

We thank Jacob Abajian, Veronica Collazuol, Scott Collier, Henry Maxfield, Dalimil Mazac, Sridip Pal, Yiannis Tsiares, and Pierfrancesco Urbani for helpful discussions. EP and GD thank the Kavli Institute for Theoretical Physics, Santa Barbara for support during the course of this work. EP also thanks the ICTP Trieste and Kavli IPMU for hospitality. GD also thanks the ICISE in Quy Nhon, Vietnam for hospitality. This research was supported by ERC Starting Grant 853507, and in part by the National Science Foundation under Grant No. NSF PHY-1748958.

\begin{appendix}

\sec{Density of states}\label{app:mwk}

We write the total spin-$j$ density of states as 
\e{eq:rhototapp}{\rho_j(t)=\rho_{{\rm MWK},\,j}(t)+ \rho_{{\rm string},\,j}(t)}
The MWK density of states may be written as \cite{Benjamin:2019stq,Benjamin:2020mfz}
\es{eq:rhomwk}{ 
&\rho_{{\rm MWK},j}(t)=\\&\frac{2}{\sqrt{t\tbar}}\sum_{s=1}^{\infty} \bigg[f_{j,0;s}\cosh(\frac{4\pi}{s}\sqrt{\xi\tbar})\cosh(\frac{4\pi}{s}\sqrt{\xi t}) \\& -f_{j,-1;s}\cosh(\frac{4\pi}{s}\sqrt{\xi\tbar})\cosh(\frac{4\pi}{s}\sqrt{(\xi-1) t})\\& -f_{j,1;s}\cosh(\frac{4\pi}{s}\sqrt{(\xi-1)\tbar})\cosh(\frac{4\pi}{s}\sqrt{\xi t})\\& +f_{j,0;s}\cosh(\frac{4\pi}{s}\sqrt{(\xi-1)\tbar})\cosh(\frac{4\pi}{s}\sqrt{(\xi-1) t})\bigg]
}
where $f_{j,k;s} : = {S(j,k;s)/s}$ and $S(j,k;s)$ is a Kloosterman sum,
\e{kldef}{S(j,k;s) = \sum_{0\leq d \leq s-1,\,(d,s)=1} e^{2\pi i {d\o s}j + {d^{-1}\o s} k}}
where $d^{-1}\in\Z$ is the multiplicative inverse of $d$ mod $s$. The scalar density requires regularization \c{Maloney:2007ud,Keller:2014xba}. It is comprised of a delta function piece given by  \eqr{eq:MWKneg2}, and a continuous piece which we denote by $\tilde{\rho}_{{\rm MWK},0}(t)$:
\es{eq:rhomwkreg}{
&\tilde{\rho}_{{\rm MWK},0}(t)=\\& \frac{2}{t} \sum_{s=1}^\i\bigg\{  {\phi(s)\o s} \bigg[\sinh^2\(\frac{4\pi}{s}\sqrt{\xi t}\)+\sinh^2\(\frac{4\pi}{s}\sqrt{(\xi-1) t}\) \bigg] \\ & -2\,{\mu(s)\o s}\bigg[\cosh(\frac{4\pi}{s}\sqrt{\xi t})\cosh(\frac{4\pi}{s}\sqrt{(\xi-1) t})-1\bigg]\bigg\}
}
where $\phi(s)=S(0,0;s)$ is the Euler totient function and $\mu(s)=S(0,1;s)$ is the Mobius function.  

The string density for $j\neq 0$ was given in \eqr{eq:rhostring}, which we repeat here for convenience:
\es{eq:rhostring:app}{\rho_{{\rm string},\,j}(t) &= \frac{4}{\sqrt{t\,\tbar}}\sum_{s=1}^{\infty} f_{j,j_*;s}\cos\Big(\frac{2\pi}{s}\sqrt{\xi \tbar}\Big)\cosh\Big(\frac{2\pi}{s}\sqrt{\xi t}\Big)\\&+(j_* \rar -j_*, t \leftrightarrow \tbar)}
Similarly to the MWK case, the scalar density requires regularization. The regularized density contains a delta function piece given in \eqr{rhoscal}, and a continuous piece which we denote by $\tilde{\rho}_{{\rm string},0}(t)$:
\es{eq:rhostringreg}{
\tilde{\rho}_{{\rm string},0}(t)= &\frac{8}{t} \sum_{s=1}^\i\frac{ c_s(j_*) }{s}\x\\&\bigg[ \cos(\frac{2\pi}{s}\sqrt{\xi t})\cosh(\frac{2\pi}{s}\sqrt{\xi t})-1\bigg]
}
where $c_s(j_*)$ is a Ramanujan sum,
\e{}{c_s(j_*) = \sum_{1\leq d \leq s,\,(d,s)=1}e^{2\pi i {d\o s} j_*}} 
The total regularized scalar density is given by the sum of these two contributions, together with the delta functions in \eqr{rhoscal}:
\e{rhoscal:app}{\rho_0(t)=\delta(t+\xi) + (-6+8\sigma_0(j_*))\delta(t) + \tilde{\rho}_{0}(t)}
where
\es{eq:regtot}{ 
\tilde{\rho}_0(t)=\tilde{\rho}_{{\rm MWK},0}(t)+\tilde{\rho}_{{\rm string},0}(t).
}

\sec{Positivity}\label{App:positivity}
We treat in turn the positivity of the large spin $j\rightarrow \infty$,  finite spin  $j\geq 1$, and scalar $j=0$ sectors of the density \eqr{eq:rhototapp}, focusing mostly on the regime $\xi \gg1$. Actually, we consider a more general case in which we have $d_*$ string states: namely, $Z_p(\t) = Z_{\rm MWK}(\t)+{d_*\o 2}\,Z_{\rm string}(\t)$, and correspondingly for the densities. For the partition function $\cZ(\t)$ defined in the main text, $d_*=2$.  

\ssec{Positivity at large spin}
The MWK density \eqr{eq:rhomwk} is known to be negative in the extremal limit $t\rightarrow 0$ of large spin $|j|\rightarrow \infty$ (more precisely, for $t\lesssim t_0 \sim e^{-2\pi \sqrt{\xi j}}$, dropping a numerical prefactor). We cure this negativity with the states $\O_*$ by design, having chosen their twist to be $t_*=-\frac{\xi}{4}$. The mechanism is the same as described in Sec. 2.1 of \cite{Benjamin:2020mfz}, building on \c{Alday:2019vdr,Benjamin:2019stq}. In the regime $t<t_0$ with $j\rightarrow \infty$, the MWK density is approximately equal to \cite{Benjamin:2020mfz}
\es{mwkextapprox}{ 
\rho_{\MWK,\,j}(t)\approx \frac{(-1)^{j}}{\sqrt{jt}} \qty(e^{2\pi \sqrt{\xi j}}+ e^{2\pi \sqrt{(\xi-1) j}})
}
In the same regime, the string density \eqr{eq:rhostring:app} has the same exponential behavior but with a positive coefficient, coming from the state of spin $-j_*$: 
\es{}{
\rho_{\st,\,j}(t)\approx \frac{d_*}{\sqrt{jt}} e^{2\pi \sqrt{\xi j}}.
}
The string density cancels the odd-spin negativity for $d_*>1$. Whereas at $d_*=1$ there are subleading negativities to take care of, requiring the addition of higher-twist states \cite{Benjamin:2020mfz}, choosing $d_*=1+\delta$ for any finite $\delta$ gives a positive extremal density. In the construction of $\cZ(\t)$ in \eqr{zmain}, we chose $d_*=2$, the smallest integer degeneracy which guarantees positivity, as a matter of naturalness. The above discussion applies equally to the regime of large negative spin $j\rightarrow -\infty$, whereupon the negativity is cured by the state $\O_*$ with spin $+j_*$.

\ssec{Positivity at $j\geq1$}
We now consider finite spin $j\geq1$. We also take $\xi \gg 1$. There are two regimes of twist $t$ to consider: the extremal limit $t\rar 0$, and fixed $t$. 

In the extremal limit, the arguments given just above are again sufficient to guarantee positivity. In particular, we have $\xi j \gg1$ in the present regime of interest; one may confirm upon inspection that the $\xi$- and $j$-dependence of $\rho_{\MWK,\,j}(t)$ and $\rho_{\st,\,j}(t)$ are such that at $\xi j \gg1$, even for finite $j$, the result of the previous subsection carries through. 

Now we consider the regime of fixed $t$. Since $\xi j\gg 1$, terms of the form $\cosh(\frac{4\pi}{s}\sqrt{\xi(t+ j)})$ for $s\geq 2$ are exponentially suppressed with respect to the $s=1$ term. At fixed $t$, the MWK density is therefore well-approximated by the $s=1$ term:
\es{mwkfinitej}{ 
\frac{\sqrt{t\tbar}}{2}&\rho_{\MWK,\,j}(t)\approx  \qty[\cosh(4\pi\sqrt{\xi t})-  \cosh(4\pi\sqrt{(\xi-1)t}) ]  \\&\times \qty[\cosh(4\pi\sqrt{\xi(t+j)})-  \cosh(4\pi\sqrt{(\xi-1)(t+j)}) ]
}
This is manifestly positive, and scales as $\sim e^{4\pi\sqrt{\xi j}}$ times an $\O(1)$ coefficient. The string density at leading order in $\xi j \gg 1$ is 
\es{eq:stringdenapp}{
\frac{\sqrt{t\tbar}}{2 d_*}&\rho_{\st,\,j}(t)= \cos(2\pi \sqrt{\xi t})\cosh(2\pi\sqrt{\xi(t+j)}) \\ &+ \sum_{s=1}^{\infty}f_{j,j_*;s}\cos(\frac{2\pi}{s}\sqrt{\xi (t+j)})\cosh(\frac{2\pi}{s}\sqrt{\xi t})
}
In the first line we dropped the exponentially-suppressed $s>1$ terms (this is allowed because the sum over $s$ cannot lead to exponential enhancement), whereas no such suppression is present in the second line. Noting that \eqr{eq:stringdenapp} scales as $\sim e^{2\pi \sqrt{\xi j}}$, we see that the sum of \eqr{mwkfinitej} and \eqr{eq:stringdenapp} is positive, as the latter is exponentially suppressed in $\xi j \gg 1$. As an aside, note that this hierarchy can be overcome if $d_*$ is exponentially large in $\xi$, a possibility that we discard (in the next subsection we bound $d_*$ by an $\O(1)$ number). 

Summarizing so far, we have shown that $\rho_j(t)>0$ for $j\geq 1$ and all $t$ at $\xi \gg 1$. 

\ssec{Positivity at $j=0$}
The scalar sector requires slightly more attention since there is no longer a parametrically large scale that suppresses $s>1$ terms in the density. Since the MWK density is exponentially large and positive as $\xi t \gg 1$, any possible negativity will arise only for $\xi t \lesssim \cO(1)$.  We can then study the density at fixed $x:= 2\pi \sqrt{\xi t}$, where we also take $\xi \gg 1$. 

We divide the proof into two parts: showing that the sum of $(s=1)$ and $(s=2)$ terms is positive for $d_*$ below a critical value; and showing that $s>2$ terms are individually positive.

\begin{figure}[t]
\centering
{
\includegraphics[scale=0.67]{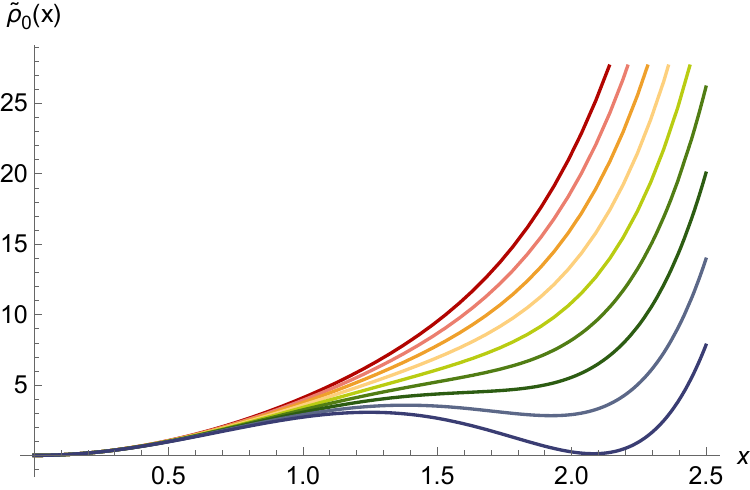}
}
\caption{Plot of the regularized scalar density of states $\tilde{\rho}_{0}(x)$ with $\xi=1000$, as a function of $x=2\pi \sqrt{\xi t}$, with degeneracy ranging from $d_*=3.3$ (red) to $d_*=7.3$ (blue) in half-integer steps. For $d_* \gtrsim 7.3$, the density develops a negative region. (Obtained by summing over $s \leq 200$ in \eqr{eq:regtot}.)}
\vspace{.1in}
\label{fig:den73}
\end{figure} 

\begin{figure}[t]
\centering
{
\includegraphics[scale=0.67]{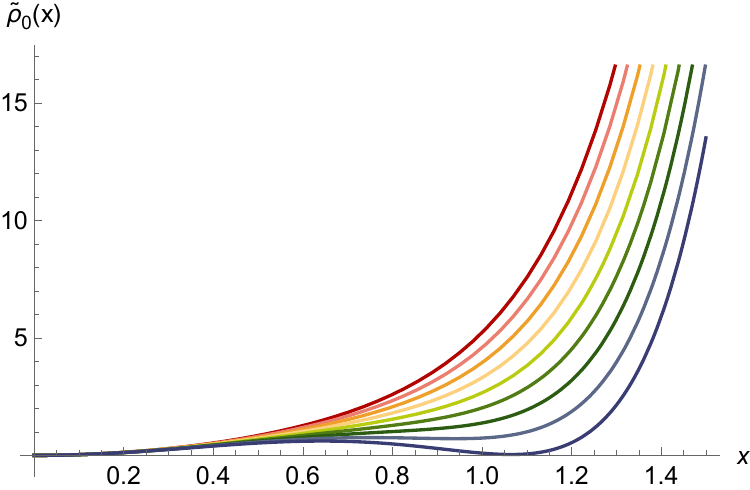}
}
\caption{Plot of the regularized scalar density of states $\tilde{\rho}_{0}(x)$  with $\xi=2$, as a function of $x=2\pi \sqrt{\xi t}$, with degeneracy ranging from $d_*=3.2$ (red) to $d_*=19.2$ (blue) in steps of two. For $d_* \gtrsim 19.2$, the density develops a negative region -- a larger critical value than for $\xi=1000$. (Obtained by summing over $s \leq 200$ in \eqr{eq:regtot}.)}
\vspace{.1in}
\label{fig:den19}
\end{figure}

\sssec{$(s=1)+(s=2)$}
The sum of the $s=1,2$ terms of \eqr{eq:rhomwkreg} and \eqr{eq:rhostringreg} (times ${d_*\o 2}$) is, at leading order in large $\xi$, 
\es{eq:rhos1s2}{ 
\frac{t}{2}\tilde{\rho}_{0}(t)\big|_{s\leq 2}&= 2\sinh^2 x+2d_* (\cos x\cosh x-1)\\&+d_*(-1)^{j_*}\qty(\cos(\frac{x}{2})\cosh(\frac{x}{2})-1).
} 
One can easily see numerically that upon increasing $d_*$, this function develops a minimum $x_{\rm min}$ which eventually becomes negative, approximately given by
\es{}{ 
d_*&\lesssim 4.910 , \quad x_{\rm min}\approx1.851 \qquad (j_* ~\text{even})\\
d_*&\lesssim 5.236 , \quad  x_{\rm min}\approx1.847 \qquad (j_* ~ \text{odd})
}
If $d_*$ obeys these bounds, then \eqr{eq:rhos1s2} is positive. We can check how these bounds compare to the full sum over $s$ at finite but large central charge: see Fig.  \ref{fig:den73}. Summing up to $s=200$ for $\xi=1000$, which easily ensures convergence, we observe numerically that the density becomes negative for $d_*\gtrsim 7.3$, not far from the limited analytic bound obtained above. The growth of the upper bound as we include more terms in the sum is due to the positivity of the $s>2$ terms, as we will show next. Note that at smaller $\xi$, the upper bound actually grows, as seen in Fig. \ref{fig:den19}: for the smallest value $\xi=2$ allowed within our construction, we observe positivity for $d_*\lesssim 19.2$.

\sssec{$s\geq3$ terms}
At leading order in $\xi \gg 1$, the density of states for $s\geq 3$ is:
\es{b5}{ 
{t s \o 2}\tilde{\rho}_{0}(t)\big|_{s\geq 3}&=  2(\phi(s)-\mu(s))\sinh^2(2x_s)\\&+2d_* c_s(j_*)\qty(\cos x_s\cosh x_s-1),
}
where 
\e{}{x_s:=\frac{x}{s}\,.}
Denoting the right-hand side of \eqr{b5} as $f(x_s)$,  we observe that there is a minimum at $x_s=0$ for which $f(0)=f'(0)=0$. As a consequence, one way to ensure positivity is to demand convexity, $f''(x_s)>0$, for all $x_s>0$; this is of course not a necessary condition, but it is sufficient to achieve our goal of demonstrating existence of a range of $d_*$ in which these terms are positive. Imposing convexity gives the inequality
\es{ineq}{ 
16(\phi(s)-\mu(s))\cosh(4x_s)-4d_* c_s(j_*)\sin x_s\sinh x_s>0\,.
}
Using the bounds 
\e{}{|c_s(j_*)|<\phi(s)\,,\quad |\mu(s)|\leq 1\,,\quad \phi(s\geq 3)\geq 2}
gives rise to the strongest inequality,
\e{}{d_* < 2{\cosh(4x_s) \o \sinh(x_s)}\,.}
If this is satisfied for all $x_s$, then so is \eqr{ineq}. Minimizing the right-hand side gives
\e{}{d_* \lesssim 11.888\,.}
This ensures positivity of each individual $s\geq 3$ term in the density. This is compatible with the previously derived bounds for the $s=1,2$ terms. 

Altogether, we conclude that for a finite range of $d_*>1$, the density of states is positive, $\rho_j(t)>0$, for all spins $j$ and twists $t$ at $\xi \gg1$.

\sec{Spectral decomposition of $Z_{\rm string}(\t)$}\label{app:zspec}
In this appendix we derive \eqr{zstspec}. We directly present the relevant calculations, directing the reader to \cite{Terras_2013,Benjamin:2021ygh, Collier:2022emf} for details on the $\sl$ spectral formalism, and \cite{Benjamin:2022pnx,Haehl:2023tkr,DiUbaldo:2023qli} for its further use in the 2d CFT context.

We wish to compute the Petersson inner product
\e{}{(Z_{\rm string},\psi_{\w}) := \int_{\mathcal{F}} {dx dy\o y^2} Z_\st(\t) \overline\psi_\w(\t)}
where $\psi_{\w}(\tau)= \{ E_{\half+i\w}(\tau),\phi_n(\tau)\}$ are the $\sl$ eigenbasis elements. Since $Z_{\rm string}(\t)$ is a Poincar\'e sum, the overlaps with the Eisenstein series and Maass cusp forms can be easily computed using the ``unfolding trick.'' This results in the following integral for the Eisenstein series:

\es{}{(Z_{\rm string},E_{\half+i\w})= \frac{4\mathsf{a}_{j_*}^{(\half+i\w)}}{\Lambda\qty(\half-i\w)}\int_0^{\infty} \frac{dy}{y} K_{i\omega} (2\pi j_* y)
}
where $\mathsf{a}_j^{(\half+i\w)}$ are the Eisenstein Fourier coefficients
\e{adef}{\mathsf{a}_j^{(\half+i\w)}= {2 \sigma_{2i\w}(j)\o j^{i\w}}\,, }
which obey reflection symmetry, $\mathsf{a}_j^{(\half+i\w)} = \mathsf{a}_j^{(\half-i\w)}$. 
The cusp form overlap is obtained similarly: 
\es{cuspoverlap}{(Z_{\rm string},\phi_n)=4 \mathsf{b}_j^{(n)} \int_0^{\infty} \frac{dy}{y} K_{i\omega} (2\pi j_* y),
}
where 
 $\mathsf{b}_j^{(n)}$ are the cusp form Fourier coefficients, known only approximately via numerics \c{lmfdb}. The integral is divergent at the origin. Regularizing the divergence is straightforward: introducing
\es{}{ 
(Z_{\rm string},E_{\half+i\w})&_{\eps}:=\frac{4\mathsf{a}_{j_*}^{\qty(\half+i\w)}}{\Lambda\qty(\half-i\w)} \int_0^{\infty} \frac{dy}{y^{1-\eps}} K_{i\omega}(2\pi j_* y)\\ &= \frac{(\pi j_*)^{-\eps}\mathsf{a}_{j_*}^{\qty(\half+i\w)}}{\Lambda\qty(\half-i\w)}   \Gamma\qty(\frac{\eps-i\w}{2})\Gamma\qty(\frac{\eps+i\w}{2})
} 
the overlaps may be defined by removing the regulator,
\es{}{ 
(Z_{\rm string}E_{\half+i\w})&= \lim_{\eps \rightarrow 0} (Z_{\rm string},E_{\half+i\w})_{\eps} \\&= \frac{\mathsf{a}_{j_*}^{\qty(\half+i\w)}}{\Lambda\qty(\half-i\w)}  \Gamma\qty(-\frac{i\w}{2})\Gamma\qty(\frac{i\w}{2}).
}
and likewise for the cusp form overlap \eqr{cuspoverlap}. This yields the spectral decomposition \eqr{zstspec}. We note that the regularization used here is equivalent to the following standard regularization of Poincar\'e sums over seed primaries of fixed dimensions,
\es{}{
Z_{h,\hb}^{\eps}(\tau) := \sum_{\gamma \in \sl/\Gamma_{\i} } \text{Im}(\gamma\tau)^{\half+\eps} q_{\gamma}^{h-\xi}\qb_{\gamma}^{\hb-\xi}\,,
 }
where $(h,\hb) = ({5\o4}\xi,{3\o4}\xi)$ for our state $\O_*$. 

\ssec{Re-deriving the scalar density}
As a consistency check, we can re-derive the scalar density $\rho_{\st,0}(t)$ from the spectral decomposition. The scalar piece of the regularized partition function is
\e{Zst0int}{\frac{Z_{\rm string,0}^{\eps}(y)}{2\sqrt{y}(\pi j_*)^{-\eps}}=  \int_{\ccrit}  \frac{y^{i\w}\mathsf{a}_{j_*}^{(\half+i\w)}}{\Lambda\qty(\half-i\w)} \Gamma\qty(\frac{\eps+i\w}{2})\Gamma\qty(\frac{\eps-i\w}{2})}
where $\int_{\ccrit}  = {1\o 4\pi} \int_{-\i}^\i d\w$ is the integration along the critical line. In writing \eqr{Zst0int} we have used the scalar Fourier modes $E^*_{s,0}(y) = \L(s) y^s + \L(1-s) y^{1-s}$ and $\phi_{n,0}(y) = 0$. We now perform contour integration for complex $z:=i\w$ by deforming to a new contour, $\mathcal{C}$, a semicircle in the left half plane  $\Re(z)<0$ such that $y^z$ decays at infinity. The integrand vanishes factorially on the arc at infinity due to the $\Lambda\qty(\half-i\w)$ in the denominator. The  poles inside the contour come from $\Gamma\qty(\frac{\eps+z}{2})$ at $z=-2k-\eps$ with $k=0,1,\dots$. The integral \eqr{Zst0int} is then given as a sum over residues,
\es{Zstreg}{ 
&\frac{Z_{\rm string,0}^{\eps}(y)}{\sqrt{\pi}(\pi j_*)^{-\eps}}
= \text{($k=0$ term) } +\\ &\sqrt{y} \sum_{k=1}^{\infty}\frac{(-1)^k}{k!}\frac{4\sigma_{4k+2\eps}(j_*)\Gamma(k+\eps)}{\Gamma(\half+\eps+2k)\zeta(1+2\eps+4k)}\qty(\frac{\pi}{j_*y})^{2k+\eps}
}
where we used $\Lambda(s)=\pi^{-s}\Gamma(s)\zeta(2s)$ and the explicit Fourier coefficients \eqr{adef}. We have separated the $k=0$ term because as we remove the regulator $\eps\rightarrow0$, two simple poles at $z=\pm \eps$ coalesce into a double pole at $z=0$, on the contour. We will thus treat separately the $k>0$ terms, for which the regulator can be trivially removed, and the $z=\pm \eps$ poles. 

Let us now Laplace transform to the density of states,
\e{}{Z_{\st,0}(y)= \sqrt{y} \int_0^{\infty} dt\, e^{-4\pi yt} \rho_{\st,0}(t).
}
We have written the density in terms of the reduced twist, $t=\D/2-\xi$ for scalars,  which is related to the density as a function of dimension $\Delta$ through $\rho(\Delta)d\Delta=\rho(t)dt$. The regularized partition function \eqr{Zstreg} gives a regularized density 
\es{}{ 
&\rho_{\rm string,0}(t)= \text{($k=0$ term) } + \\& {8 \sqrt{\pi}\o t}\sum_{k=1}^{\infty}\frac{(-1)^k\sigma_{4k}(j_*)}{\Gamma(1+2k)\Gamma(\half+2k)\zeta(1+4k)}\qty(\frac{4\pi^2 t}{j_*})^{2k}
}
where we have removed the regulator in the second line. Using the identity
\e{}{\sigma_z(j)=\zeta(z+1)j^z\sum_{s=1}^\i\frac{c_s(j)}{s^{z+1}}\,,}
swapping the order of the sums and performing some simplifications,
\es{}{
\rho_{\rm string,0}(t) &=   \text{($k=0$ term) } + \\&\frac{8}{t}\sum_{s=1}^\i\frac{c_s(j_*)}{s}\sum_{k=1}^\i\frac{(-1)^{k}}{(4k)!}\qty(\frac{2\pi\sqrt{2\xi t}}{s})^{4k}\,.
}
The second line can be resummed to reproduce \eqr{eq:rhostringreg}, the continuous part of the scalar density.

Finally, we return to the $k=0$ term, still with the regulator, which is explicitly given by
\es{}{ 
\rho_{\rm string,0}(t)\big|_{k=0}=\frac{4 \sqrt{\pi}}{t(\pi j_*)^{\eps} } \frac{\sigma_{2\eps}(j_*)}{\Gamma(\half+\eps)\zeta(1+2\eps)}\qty(\frac{4\pi^2 t}{j_*})^{\eps}
}
To regulate the divergence as $\eps\rightarrow 0$ and $t\rar 0$, similarly to \cite{Benjamin:2020mfz} we integrate from $t=0$ up to some $t_*>0$ and use $\eps\zeta(1+2\eps)\rar\half$ as $\eps\rar 0$ to arrive at 
\e{}{ 
\lim_{\eps\rightarrow0}\left[\int_0^{t_*} dt \rho_{\rm string,0}(t)\big|_{k=0}\right]= 8\sigma_0(j_*). 
}
Together with the continuous part of the density derived above, this reproduces the full result \eqr{rhoscal:app}.

\end{appendix}

\bibliography{stringdraftbib.bib}

\end{document}